\documentclass[aps, pra, 10pt, twocolumn, superscriptaddress, floatfix,longbibliography]{revtex4-1}
\usepackage[charter]{mathdesign}
\usepackage{amsmath,graphicx,bm}
\usepackage{tikz}
\usepackage[caption=false]{subfig}
\usepackage[pdftex,plainpages=false,colorlinks=true,linkcolor=blue, citecolor=blue, urlcolor=blue]{hyperref}
\newcommand{\cud}{$3d_{x^2-y^2}$~}
\newcommand{\e}{\epsilon}

\newcommand{\dw}{\downarrow}
\newcommand{\etal}{\textit{et al.}}

\newcommand\kv{\mathbf{k}}

\newcommand\tv{\mathbf{t}}

\newcommand\kvt{\mathbf{\tilde k}}

\newcommand\Gv{\mathbf{G}}

\newcommand\Sigmav{\bm{\Sigma}}

\newcommand\s{\sigma}

\newcommand\up{\uparrow}

\newcommand\Gammav{\bm{\Gamma}}

\newcommand\om{\omega}

\begin{document}
\title{Pseudogap transition within the superconducting phase in the three-band Hubbard model}

\author{S. S. Dash}\affiliation{D\'epartement de physique and Institut quantique, Universit\'e de Sherbrooke, Sherbrooke, Qu\'ebec, Canada J1K 2R1}
\author{D. S\'en\'echal}\affiliation{D\'epartement de physique and Institut quantique, Universit\'e de Sherbrooke, Sherbrooke, Qu\'ebec, Canada J1K 2R1} 
\date{\today}

\begin{abstract}
	The onset of the pseudogap in high-$T_c$ superconducting cuprates (HTSC) is 
	marked by the $T^*$ line in the doping-temperature phase diagram, which ends 
	at a point $p^*$ at zero temperature within the superconducting dome. 
	Although various theoretical and experimental studies indicate a competition 
	between the pseudogap and superconductivity, there is no 
	general consensus on the effects of the pseudogap within the superconducting 
	phase. We use cluster dynamical mean field theory on a three-band Hubbard 
	model for the HTSC to study the superconducting phase at $T=0$, obtained 
	when doping the charge-transfer insulator, for several values of $U$. We observe 
	a first-order transition within the superconducting phase, which separates 
	the underdoped and overdoped solutions. The transition to the 
	underdoped solution is marked by a discontinuous increase in the spectral 
	gap, and on further underdoping the spectral gap increases while the 
	superconducting order parameter decreases. We conclude that this is due 
	to the onset of the pseudogap in the underdoped region, which contributes 
	to the increasing spectral gap; this is further consistent with the 
	appearance of a pole in the normal component of the self-energy,
	in the antinodal region, in the underdoped solution. This is accompanied by a 
	change in the source of the condensation energy from potential energy,
	in the overdoped region, to kinetic energy in the underdoped region. 
	Further, we also observe that the $d$-wave node vanishes smoothly within 
	the superconducting phase at low values of hole doping, within the 
	underdoped region. We see this as a manifestation of Mott physics 
	operating at very low doping. Various aspects of the results and 
	their implications are discussed.

\end{abstract}

%\pacs{71.27.+a, 71.10.Fd, 71.10.Hf, 71.30.+h}
\maketitle

%================================================================================
\section{Introduction}

One of the most striking effects of strong correlations in hole-doped high-$T_c$ superconducting cuprates (HTSC) is the pseudogap (PG).
It manifests itself as a loss of density of states along the antinodal directions at temperatures less than $T^*$~\cite{timusk1999pseudogap}. 
It has been considered a precursor of superconductivity, which would emerge on lowering the temperature further, 
below $T_c$~\cite{emery1995importance,renner1998pseudogap,timusk1999pseudogap}. 
However, such a picture has fallen out of favor since it has been observed that the $T^*$ line ends within the superconducting 
dome at a doping $p^*$~\cite{tallon2001doping,daou2009linear}. 
This means that the pseudogap and superconductivity have distinct origins.

Therefore, an important question to consider is whether the pseudogap coexists with $d$-wave superconductivity.
Various studies have found evidence for the pseudogap within the superconducting (SC) phase
~\cite{mcelroy2005coincidence,vishik2012phase,loret2017vertical}. 
Although the doping dependence is not very well understood, one view is that the magnitude of the gap 
does not vary much with doping, in the pseudogap phase below $T_c$~\cite{mcelroy2005coincidence,vishik2012phase,hashimoto2014energy}. 
On the other hand, Tanaka \etal~\cite{tanaka2006distinct} report that the anti-nodal gap, attributed to the 
pseudogap, increases with underdoping while the near-nodal gap, 
seen as a proxy to the SC gap, decreases. Kondo \etal~\cite{kondo2009competition} observe a 
similar competition between superconductivity and the pseudogap. Additionally, there is evidence 
of a nodeless SC gap at very low values of hole doping~\cite{vishik2012phase,razzoli2013evolution};
whether it is related to the pseudogap is not clear. 
On the theoretical side, there have been studies indicating strong signatures of a quantum critical point within the 
SC state associated with strong momentum-space differentiation~\cite{haule2007avoided,civelli2009doping}. 
In particular, Civelli~\etal~\cite{civelli2008nodal} suggest two gap energy scales within the SC phase. 
These studies point to an inherent competition between the pseudogap and superconducting phases. 
%Mallik~\etal ~\cite{mallik2018surprises} find nodeless extended $s$-wave superconductivity in the underdoped regime.  

The relevant physics of HTSC lies mostly in the copper oxide planes.
It has long been thought that the one-band Hubbard model should capture the basic physics of cuprates 
($d$-wave superconductivity and the pseudogap), but it is only since the advent of sophisticated numerical
methods that this could be confirmed~\cite{macridin2005physics,kancharla2008anomalous,corboz2011stripes,aichhorn2006variational}. 
A review on the origin of the pairing interaction in the 
Hubbard model is given in ref~\cite{scalapino2012common}. 
Cluster extensions of DMFT have been particularly successful 
in capturing the strong correlation physics in cuprates~\cite{macridin2005physics,kancharla2008anomalous}. 
These methods take into account the short-range correlations, which are 
crucial for $d$-wave superconductivity~\cite{balzer2010importance}. However, the one-band model fails to capture 
the dynamics of holes in cuprates, which mostly reside on the oxygen orbitals
~\cite{gauquelin2014atomic}.
A more accurate model is a three-band Hubbard model~\cite{emery_1987_Theory, varma_1987_charge, andersen1995lda}, 
which involves the copper \cud orbital and the 
two oxygen $2p$ orbitals in each $\rm CuO_2$ unit cell.
A recent work by Fratino~\etal~\cite{fratino2016pseudogap} using the three-band Hubbard model shows that 
holes are indeed located in the oxygen orbitals upon doping.

In this work, we study the SC state appearing when doping the charge-transfer insulator 
within the three-band Hubbard model, using cluster dynamical mean field theory (CDMFT) with an exact 
diagonalization impurity solver at zero temperature.
We observe three SC regimes in our computations: 
i) At large hole doping, BCS-like superconductivity with a symmetric gap in the density of states (DoS); 
ii) at moderate hole doping, superconductivity coexists with the pseudogap, and the DoS displays a large, asymmetric gap; 
this is separated from the first regime by a first-order transition; 
iii) at low doping, superconductivity becomes fully gapped. 
Although our observations seem to be in line with the zero-temperature scenario of the trisected SC dome with 
three distinct phases proposed by Vishik~\etal~\cite{vishik2012phase}, we observe signatures of only one 
discontinuous transition, corresponding to the onset of the pseudogap, i.e., between i) and ii).
We ignore the possibility of magnetic phases, in particular antiferromagnetism, in order to keep the computations simple.

The paper is organized as follows: In section \ref{sec_model_method}, we present the model and briefly describe the 
method used. In section \ref{sec_results}, we show our results and propose possible interpretations. 
Broader implications of our results are discussed in Sect.~\ref{sec_discussion}. 

%===============================================================================
\section{Model and method}\label{sec_model_method}

%-------------------------------------------------------------------------------
\subsection{The three-band Hubbard model}

We use a modified three-band Emery model~\cite{andersen1995lda} to describe the $\rm CuO_2$ planes, 
consisting of a Cu \cud  orbital and two O $2p$ orbitals within a unit cell.
The Hamiltonian is expressed as
\begin{equation} \label{eq:full_ham}
H = H_0 + U_d \sum_{i}n^{(d)}_{i\up}n^{(d)}_{i\dw}
%H = -\sum_{i,j,\s}t_{ij}^{(pd)}\left( d^{\da}_{i\s}p_{j\s} + \Hc \right) - 
%\sum_{i,j,\s}t_{ij}^{(pp)}\left( p^{\da}_{i\s}p_{j\s} + \Hc \right) \\
%+ (\e_p-\mu)\sum_i n_i^{(p)} + (\e_d-\mu)\sum_i n_i^{(d)} + U_d \sum_{i}n^{(d)}_{i\up}n^{(d)}_{i\dw}
\end{equation}
%Here $d^{\da}_{i\s}$ creates an electron with spin $\s$ in the \cud orbital at copper site $i$ and $p^{\da}_{i\s}$ creates an 
%electron in the $2p$ orbital located at oxygen site $i$.
%The parameters $\e _d$ and $\e _p$ are the orbital energies of the 3d and 2p orbitals, respectively.
%The hopping amplitudes $t_{ij}^{(pd)}$ (between Cu and O) and $t_{ij}^{(pp)}$ (between oxygens) have the range and sign indicated on Fig.~\ref{fig:cluster_cu_o}.
%The first-neighbor hopping integrals between $3d$ and $2p$ orbitals (brown bonds on Fig.~\ref{fig:cluster_cu_o}) have absolute value $t_{pd}$; the diagonal hopping integrals between oxygens (green bonds on Fig.~\ref{fig:cluster_cu_o}) have absolute value $t_{pp}$, and the second-neighbor hopping integrals between oxygens (orange bonds on Fig.~\ref{fig:cluster_cu_o}) are $t_{pp'}$.
$H_0$ is the non-interacting Hamiltonian. $U_d$ is the on-site Coulomb repulsion on the Cu \cud orbitals. 
$n^{(d)}_{i\sigma}$ is the number operator for electrons with spin-$\sigma$ at the copper site i.
The Coulomb repulsion $U_p$ on the O $2p$ orbitals is neglected.

%~~~~~~~~~~~~~~~~~~~~~~~~~~~~~~~~~~~~~~~~~~~~~~~~~~~~~~~~~~~~~~~~~~~~~~~~~~~~~~~
% Fig.~1
\begin{figure}
	\centering{
	\includegraphics[scale=0.65]{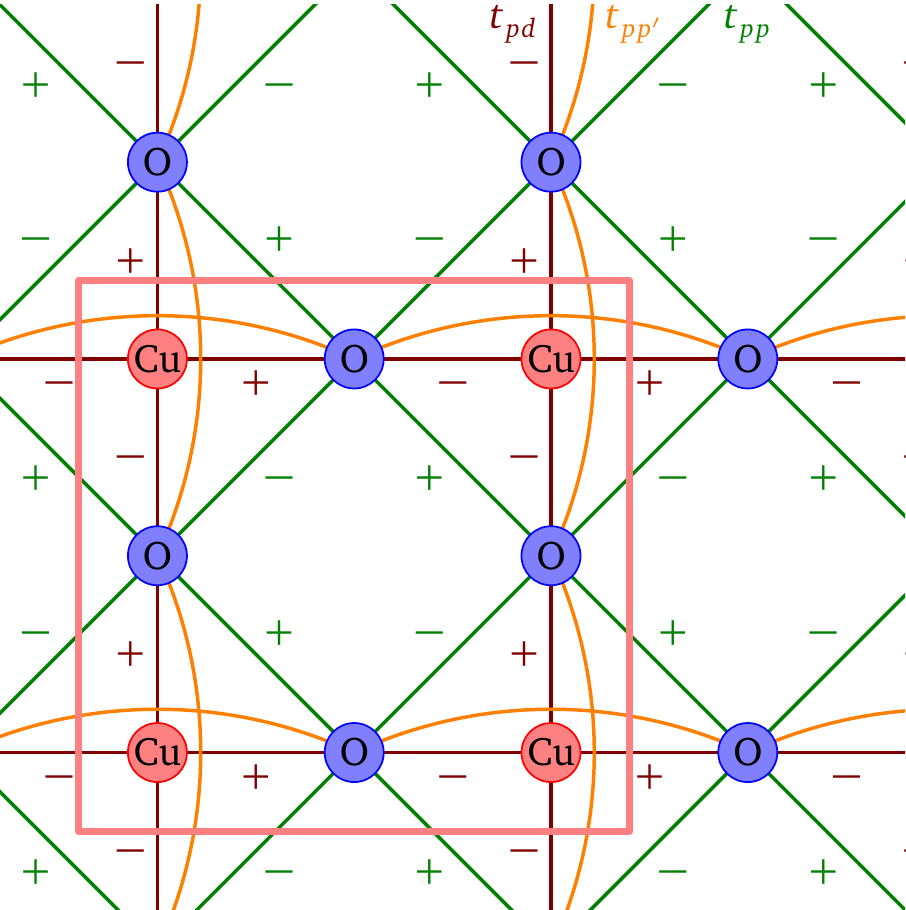}}
	\caption{Sketch of the CuO$_2$ lattice.
	The signs on the hopping terms in \eqref{eq:non-int_ham} are indicated and are the result of 
	the combined phase of the orbitals participating in bonding.
	The four-site cluster of copper atoms used in CDMFT is indicated (red box).}
	\label{fig:cluster_cu_o}
	\end{figure}
	%~~~~~~~~~~~~~~~~~~~~~~~~~~~~~~~~~~~~~~~~~~~~~~~~~~~~~~~~~~~~~~~~~~~~~~~~~~~~~~~
		
	The non-interacting Hamiltonian $H_0$ is often expressed as a hopping matrix in $\kv$-space:
\begin{widetext}
\begin{equation}
H_0(\kv) = \begin{pmatrix}
\e _d-\mu & t_{pd}(1-e^{-ik_x}) & t_{pd}(1-e^{-ik_y})\\
t_{pd}(1-e^{ik_x}) & \e _p -\mu + 2t_{pp'}\cos k_x & t_{pp}(1-e^{ik_x})(1-e^{-ik_y}) \\
t_{pd}(1-e^{ik_y}) & t_{pp}(1-e^{-ik_x})(1-e^{ik_y}) & \e _p -\mu + 2t_{pp'}\cos k_y
\end{pmatrix}\label{eq:non-int_ham}
\end{equation}
\end{widetext}
$\e _d$ and $\e _p$ are the orbital energies of the 3d and 2p orbitals respectively. $\mu$ is the chemical potential. 
$t_{pd}$ is the absolute value of the first neighbor hopping integral between $3d$ and $2p$ orbitals (brown bonds in Fig.~\ref{fig:cluster_cu_o}), 
$t_{pp}$ is the absolute value of the first neighbor hopping integral between 
two $2p$ orbitals (green bonds in Fig.~\ref{fig:cluster_cu_o}) and $t_{pp'}$ is the absolute value of the hopping integral between two 2p orbitals 
separated by a Cu atom (orange bonds in Fig.~\ref{fig:cluster_cu_o}). Note that the Cu orbitals are not directly connected.

We use two sets of parameters in this work:
One is taken from Fratino~\etal~\cite{fratino2016pseudogap} and provides a simple scenario for obtaining a 
charge-transfer gap~\cite{zaanen1985band}: 
\begin{equation}\label{eq:fratino_params}
t_{pp}= 1~,~
t_{pp'}= 1~,~
t_{pd}= 1.5~,~
\e _d = 0~,~
\e _p = 7 
\end{equation}
The second is more realistic, corresponds to Bi-2212 and is obtained from ab-initio 
calculations by Weber~\etal~\cite{weber2012scaling}: 
\begin{equation}\label{eq:weber_params}
	t_{pp}= 1~,~
	t_{pp'}= 0.2~,~
	t_{pd}= 2.1~,~
	\e _d = 0~,~
	\e _p = 2.5 
\end{equation}

%-------------------------------------------------------------------------------
\subsection{Cluster dynamical mean field theory}

In cluster dynamical mean field theory, the infinite lattice is tiled into identical clusters, and each cluster's lattice environment is replaced by a set of uncorrelated orbitals (the `bath').
This cluster-bath system defines an Anderson-impurity model, which must be solved for the Green function $\Gv_{c}(\omega)$ using an \textit{impurity solver}.
In this work, we use an exact diagonalization solver at $T=0$ and therefore are limited to a small number of bath orbitals.
The cluster self-energy $\Sigmav (\omega)$ is then extracted using Dyson's Equation 
\begin{equation}\label{eq:Dyson}
\Gv_{c}(\omega)^{-1}=\omega-\tv_c-\Gammav(\omega)-\Sigmav(\omega)
\end{equation}
where $\tv_c$ is the hopping matrix on the cluster and $\Gammav(\omega)$ the \textit{hybridization function}, which depends on the bath parameters, i.e., 
the energies of the uncorrelated orbitals and their hybridization with the cluster. Note that here we have used the symbol $\Gammav$ 
for the hybridization function, instead of $\mathbf{\Delta}$ which is generally used in the context of DMFT. The cluster self-energy is then used as an 
approximation to be the full lattice self-energy, so that the lattice Green function $\Gv(\kvt , \omega)$ is expressed as:
\begin{equation}
\Gv(\kvt , \omega)^{-1}= \om - \tv(\kvt)-\Sigmav (\omega)
\label{lattice_G}
\end{equation}
where $\kvt $, the \textit{reduced} wave vector, belongs to the Brillouin zone of the super-lattice and $\tv(\kvt)$ is the one-body matrix of the model, 
expressed in a mixed basis of cluster sites and reduced wave vector $\kvt$.
The bath parameters are chosen in such a way as to minimize the difference between $\Gv_{c}(\omega)$ and the 
Fourier transform of $\Gv(\kvt,\omega)$, i.e., its local version.
Details can be found in references~\cite{senechal2010bath,kancharla2008anomalous,senechal2012cluster,pavarini13}.

In this work, the impurity cluster (red box in Fig.~\ref{fig:cluster_cu_o}) contains 4 Cu atoms which are connected to the bath. 
As far as the CDMFT procedure is concerned, only the copper part of the lattice Green function $\Gv_{\rm Cu}(\kvt,\om)$ is taken into account:
%It is important to note that, in principle, the oxygen atoms should have been included in the impurity cluster 
%Since the oxygen atoms are uncorrelated, they can be accounted for
\begin{equation}\label{eq:lattice_G_cu}
\Gv_{\rm Cu}(\kvt,\om)^{-1} = \om - \tv_{\rm Cu}(\kvt)-\Gammav_{\rm O}(\kvt,\om)-\Sigmav(\om)
\end{equation}
which contains a fixed hybridization function $\Gammav_{\rm O}(\kvt,\om)$ coming from the oxygen orbitals.
%This can also be seen as the result of tracing over the oxygen degrees of freedom when computing the Green function.
Since the oxygen orbitals are uncorrelated in our model, their effect can be exactly represented by the hybridization function 
$\Gammav_{\rm O}(\kvt,\om)$~\cite{fratino2016pseudogap}. The presence of $\Gammav_{\rm O}(\kvt,\om)$ in the lattice Green function 
ensures that the effect of the oxygen orbitals is included in the self-energy through the CDMFT self-consistency procedure.

%The effect of the uncorrelated oxygen orbitals is included in the self-energy through the CDMFT self-consistency procedure only.

% Ideally, the impurity cluster should include the oxygen orbitals as well, but this 
% would make it too demanding for the exact diagonalization solver. 
% Hence, as an approximation, oxygen orbitals are not included in the cluster~\cite{fratino2016pseudogap}. 
% This amounts to neglecting the effect of the oxygen orbitals in the cluster self-energy, 
% through the bath orbitals. Since the oxygen orbitals are uncorrelated 
% in our model, this can be considered a good approximation.

In this work, the cluster-bath system contains 4 copper sites and 8 bath orbitals.
In order to probe superconductivity, we include anomalous hybridizations between bath and cluster, along with regular hybridizations.
The Nambu formalism is used in order to incorporate both the normal and anomalous components of the Green function into a single object.
The bath parametrization is based on the irreducible representations of the point group $C_{2v}$~\cite{foley2019coexistence,koch2008sum}. 
In this bath parametrization, the bath Hamiltonian is diagonal and each bath orbital is connected to all cluster sites.
Details of the bath parametrization using the point group $C_{2v}$ can be found in Foley~\etal~\cite{foley2019coexistence}.

The average value of an one-body operator $\hat{O}=\sum_{\alpha, \beta}O_{\alpha\beta}d^{\dagger}_{\alpha}d_{\beta}$, where
$d_{i\s}$ annihilates an electron of spin $\s$ on Copper site $i$,
is obtained from the lattice Green function \eqref{eq:lattice_G_cu} as 
\begin{equation}
	\label{lattice_average}\langle \hat{O} \rangle = \oint\frac{d\omega}{2\pi}\int\frac{d^2\tilde{\mathbf{k}}}{(2\pi)^2}\rm{tr}\left[
		\mathbf{O}(\tilde{\mathbf{k}})\mathbf{G}_{\rm Cu}(\tilde{\mathbf{k}},\omega)\right]
\end{equation}
where $\kvt $ is a reciprocal lattice vector of the super-lattice. 
% $G(\kvt , \omega)$ has the periodicity of the super-lattice but not the periodicity of the lattice because the self-energy $\Sigma(\omega)$ is extracted from the local cluster Greens function $G_c({z})$ which has no information about the lattice.

%===============================================================================
\section{Results}\label{sec_results}

\begin{figure}
\centering
\includegraphics[scale=0.5]{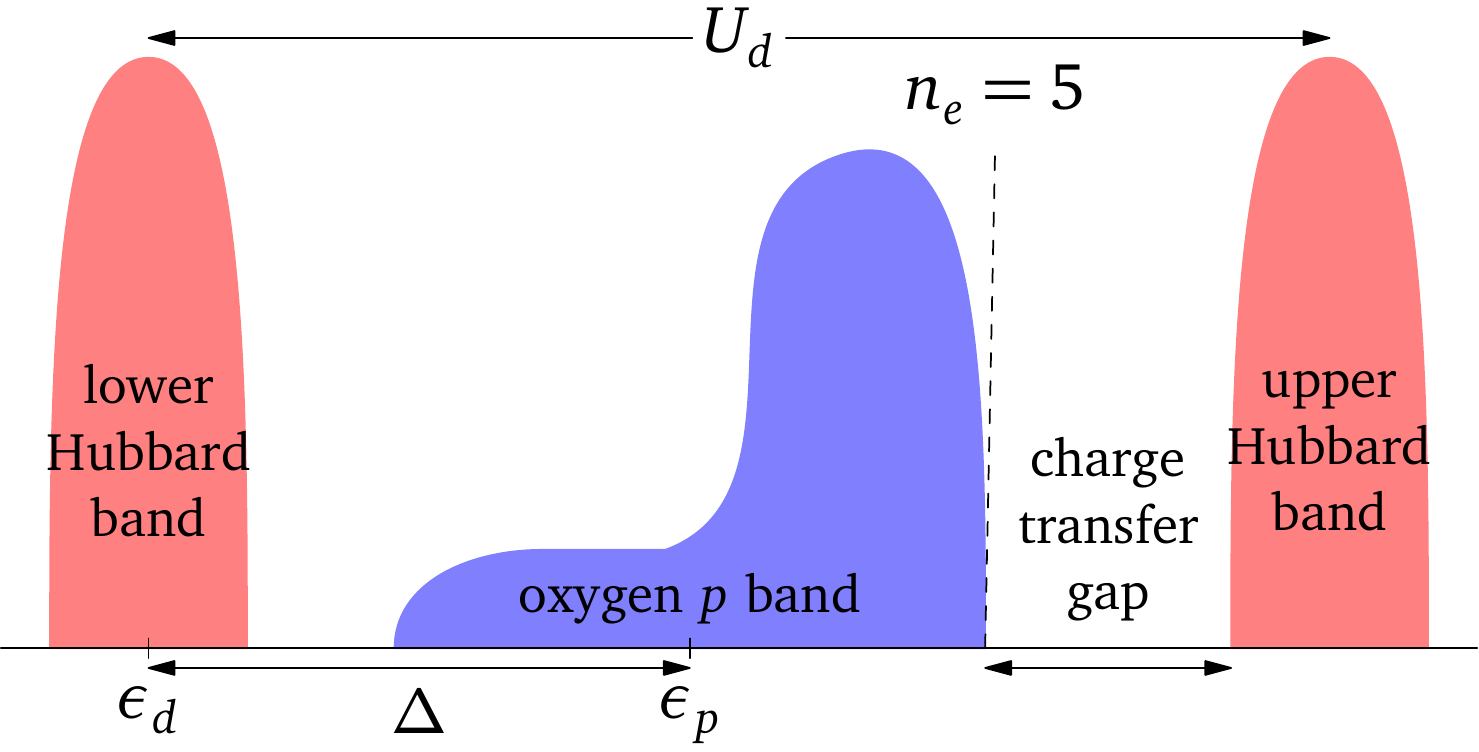}
\caption{Cartoon of the density of states (DoS) of the three-band Hubbard model.
$U_d$ splits the Copper band (red) into two subbands; the system is insulating at a filling of 5 electrons per unit-cell 
when $U_d$ is large enough to push the upper subband beyond the oxygen band.
Because electrons migrate from the Cu $3d$ orbitals to the O $2p$ as $U_d$ is increased, the insulator formed is called a 
charge-transfer insulator (CTI) and the associated gap is called the charge-transfer gap.}
\label{fig:dos_sketch}
\end{figure}

At a filling of 5 electrons per unit-cell, the system is a charge-transfer insulator (CTI). This is different from a 
Mott insulator in the sense that the insulating gap is not between the two Hubbard bands but between the oxygen band 
and the upper Hubbard band. 
On doping with holes, these primarily go into the O $2p$ orbitals, as observed in experiments~\cite{gauquelin2014atomic}.
This is consistent with our observations and is evident from the cartoon in Fig.~\ref{fig:dos_sketch}.
Within our model, this is because the orbital energy of the O $2p$ orbitals is higher than that of the Cu \cud orbitals, 
making it expensive for electrons to reside in the oxygen orbitals.

%~~~~~~~~~~~~~~~~~~~~~~~~~~~~~~~~~~~~~~~~~~~~~~~~~~~~~~~~~~~~~~~~~~~~~~~~~~~~~~~
% Fig.~2
%~~~~~~~~~~~~~~~~~~~~~~~~~~~~~~~~~~~~~~~~~~~~~~~~~~~~~~~~~~~~~~~~~~~~~~~~~~~~~~~

\begin{figure}
\centering
\includegraphics[trim=0cm 0.9cm 1.1cm 1.5cm, clip,scale=0.8]{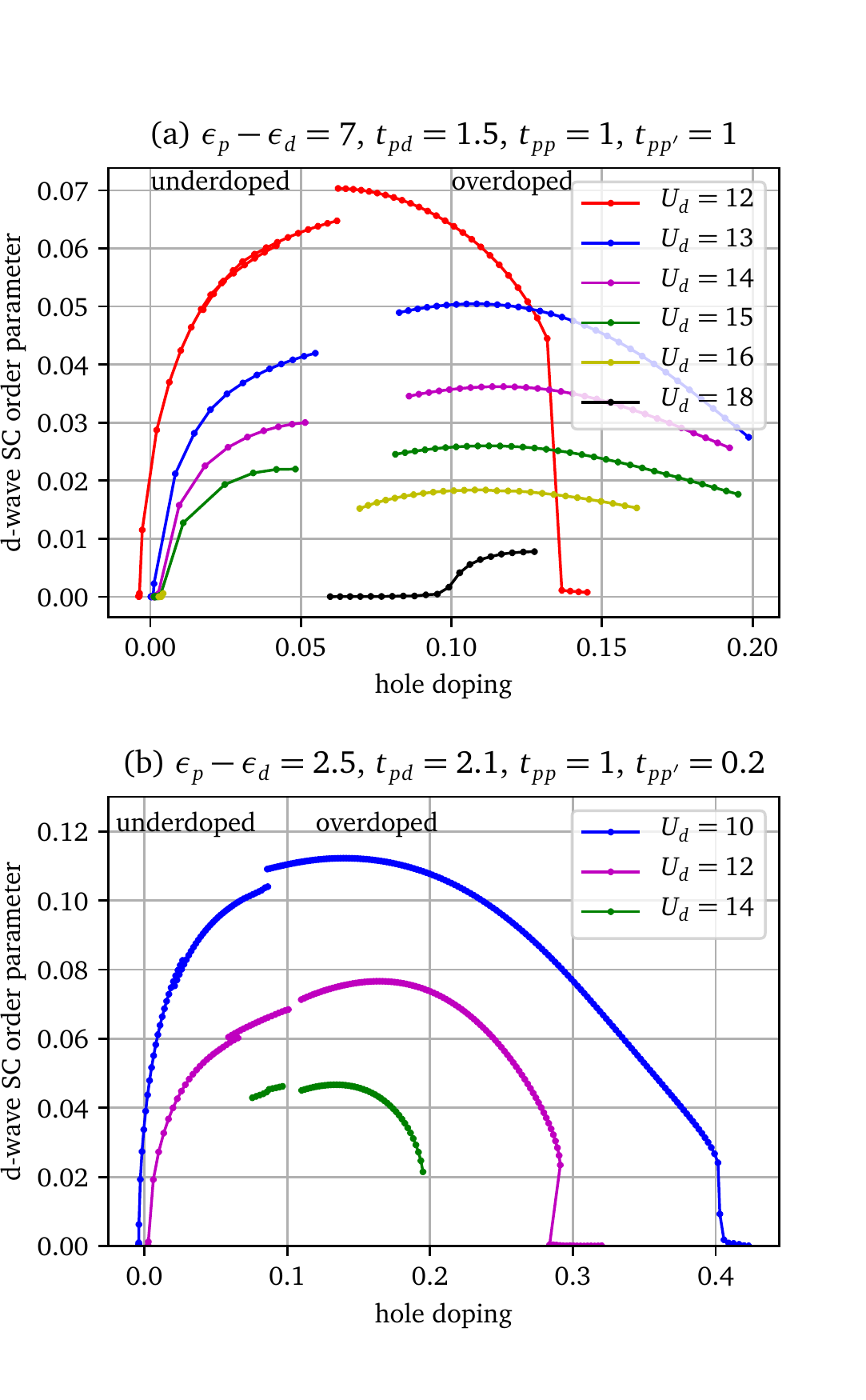} 
\caption{$d$-wave order parameter \textit{vs} doping at different values of $U_d$, 
higher than the critical value for the metal-insulator transition at a filling 
of 5 electrons in the unit-cell, for (a) Parameters \eqref{eq:fratino_params} 
and (b) Parameters \eqref{eq:weber_params}. The critical value of $U_d$ is 
around $11.7$ for Parameters \eqref{eq:fratino_params} and 
around $9.2$ for Parameters \eqref{eq:weber_params}.
}
\label{fig:d_wave_sc}
\end{figure}

Doping the CTI makes it susceptible to $d$-wave superconductivity~\cite{maier2000d}.
Figure~\ref{fig:d_wave_sc} shows the $d$-wave order parameter computed from the 
lattice Green function obtained from the converged CDMFT solutions, 
as a function of hole doping, for several values of $U_d$, all beyond the critical value $U^c_d$, that is, 
beyond the metal-insulator transition point.
The order parameter is defined as $\psi = \langle\hat\Delta\rangle/N_s$, where $N_s$ is the number of sites in the lattice and
\begin{equation}\label{eq:delta} 
	\hat\Delta = \sum_{\langle ij\rangle_x}\left(d_{i,\up}d_{j,\dw}-d_{i,\dw}d_{j,\up}\right)
	-\sum_{\langle ij\rangle_y}\left(d_{i,\up}d_{j,\dw}-d_{i,\dw}d_{j,\up}\right) + \mathrm{H.c.}
	\end{equation}
where $\langle ij\rangle_x$ indicates a sum over nearest-neighbor copper sites in the $x$ direction, and likewise for the $y$ direction.
In practice, it is computed from the anomalous part of the Green function (the Gor'kov function).

% The order parameter is computed from
% \begin{equation}
% \psi = \sum_\kvt \frac{d\omega}{2\pi} \textrm{tr}\left[\Delta(\kvt) F (\kvt,\omega)\right]
% \end{equation}
% where $\Fv(\kvt,\omega)$ is the Gorkov function and $\Delta(\kvt)$

For all values of $U_d$, except $U_d=18$ with Parameters \eqref{eq:fratino_params}, the order parameter reveals the existence 
of two solutions, which we label ``underdoped'' and ``overdoped''.
In particular, a hysteresis in the value of the order parameter is observed for $U_d=12$ and 
Parameters \eqref{eq:fratino_params} (Fig.~\ref{fig:d_wave_sc}(a)) as well as for $U_d=10$ and Parameters \eqref{eq:weber_params}
(Fig.~\ref{fig:d_wave_sc}(b)). This indicates a first-order transition between the two solutions.
For higher values of $U_d$ ($U_d>12$ for Parameters \eqref{eq:fratino_params} and $U_d>10$ for Parameters \eqref{eq:weber_params}), 
there is a range of chemical potential $\mu$ between the 
overdoped and underdoped solutions in which the CDMFT procedure does not converge, which indicates a fundamentally 
unstable region that cannot be probed with our discrete bath framework. We assume this to be fundamentally similar to 
the first-order transitions seen at lower values of $U_d$, since both situations share the 
same physics across the discontinuity as we discuss later.
The on-site Coulomb interaction $U_d$ tends to suppress the order parameter. 
Furthermore, the underdoped solution with Parameters~\eqref{eq:fratino_params} has disappeared at $U_d=18$ (Fig.~\ref{fig:d_wave_sc}(a)).

The slightly negative hole doping seen in Fig.~\ref{fig:d_wave_sc} is due 
to our use of the lattice average of electron density, as per Eq.~\eqref{lattice_average}, 
instead of the average computed from the impurity model ground state. The latter 
cannot be used since the oxygen orbitals are not contained in the impurity model.

Let us note that our system is in no way biased towards $d$-wave superconductivity except for the fact that the bath parametrization is based on the irreducible representations of the point group $C_{2v}$, which is compatible with the $d$-wave symmetry.
In principle, we could also find extended $s$-wave superconductivity, which is also compatible with the $C_{2v}$ point group, but the corresponding order parameter vanishes in our solutions.

%~~~~~~~~~~~~~~~~~~~~~~~~~~~~~~~~~~~~~~~~~~~~~~~~~~~~~~~~~~~~~~~~~~~~~~~~~~~~~~~
% Fig.~3
%~~~~~~~~~~~~~~~~~~~~~~~~~~~~~~~~~~~~~~~~~~~~~~~~~~~~~~~~~~~~~~~~~~~~~~~~~~~~~~~

\begin{figure}[htbp!]
\centering
\includegraphics[trim=0.4cm 0.45cm 0.4cm 0.4cm, clip, width=0.47\textwidth]{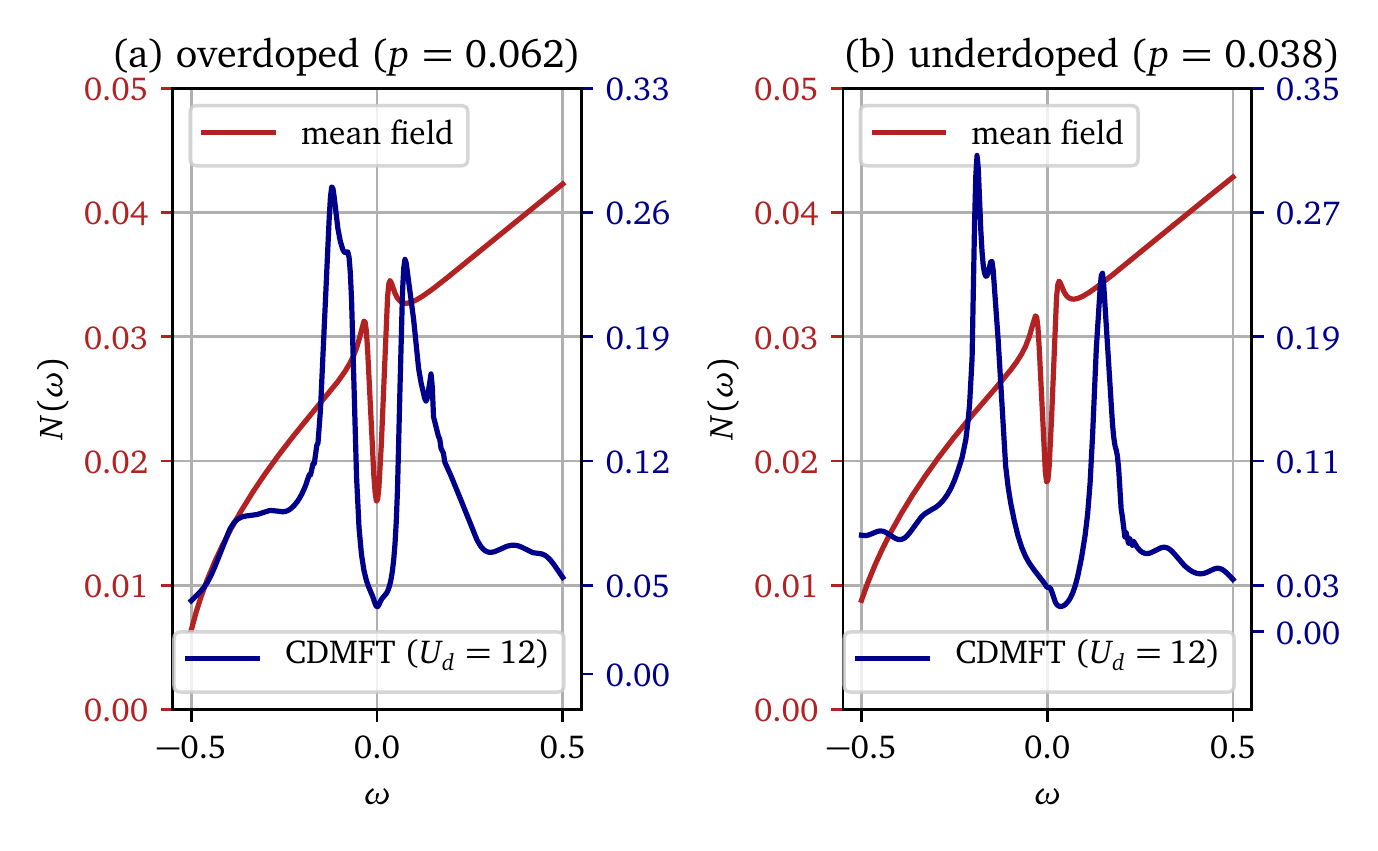}
\caption{The DoS of the CDMFT solution for Parameters \eqref{eq:fratino_params} at $U_d=12$ (in blue) 
compared with the mean-field DoS (in red), for both (a) overdoped and (b)  underdoped solutions.
The parameters of the mean-field Hamiltonian are adjusted to yield the same order 
parameter and density as those of the corresponding CDMFT solution. 
Note that the background of the gap in the CDMFT solutions is entirely different than that in the mean-field solutions, 
indicating a non-trivial redistribution of quasiparticle weight compared to the uncorrelated dispersion.
}
\label{fig:DoS_uncor_compare}
\end{figure}

The onset of superconductivity opens up a $d$-wave gap in the spectrum~\cite{wells1992evidence,hardy1993precision}, which results in a partial gapping out of the DoS at low energy.
We can get some insight into the nature of the underdoped and overdoped solutions by looking at the DoS close to the Fermi energy (Fig.~\ref{fig:DoS_uncor_compare}).
The superconducting (SC) gap, both in the underdoped and overdoped CDMFT solutions, is compared to the SC gap within a mean-field model with Hamiltonian $H_{\rm MF}=H_0 + \Delta\hat\Delta$.
The $d$-wave mean-field $\Delta$ and the chemical potential in $H_0$ are adjusted so that the order parameter and the electron density match the corresponding CDMFT solution.

The mean-field DoS obtained from $H_{\rm MF}$ (red curves in Fig.~\ref{fig:DoS_uncor_compare}) contains the pure $d$-wave gap, in contrast with the SC gap arising from strong correlation effects in our CDMFT solutions (blue curves).
In the overdoped solution (Fig.~\ref{fig:DoS_uncor_compare}(a)), the SC gap is qualitatively similar in shape to that of the pure $d$-wave mean-field SC gap.
However, in the underdoped solution (Fig.~\ref{fig:DoS_uncor_compare}(b)), the gap is strikingly different from the corresponding mean-field gap as well as from the gap in the overdoped solution: It is asymmetric and noticeably wider.
This reveals the non-trivial effects of strong correlations in the underdoped solution. Hence, there is a fundamental difference in the nature of the underdoped and overdoped solutions.

%~~~~~~~~~~~~~~~~~~~~~~~~~~~~~~~~~~~~~~~~~~~~~~~~~~~~~~~~~~~~~~~~~~~~~~~~~~~~~~~
% Fig.~4
%~~~~~~~~~~~~~~~~~~~~~~~~~~~~~~~~~~~~~~~~~~~~~~~~~~~~~~~~~~~~~~~~~~~~~~~~~~~~~~~

\begin{figure}
\centering
\includegraphics[trim=0.5cm 0.0cm 0.0cm 1.0cm, clip,scale=0.7]{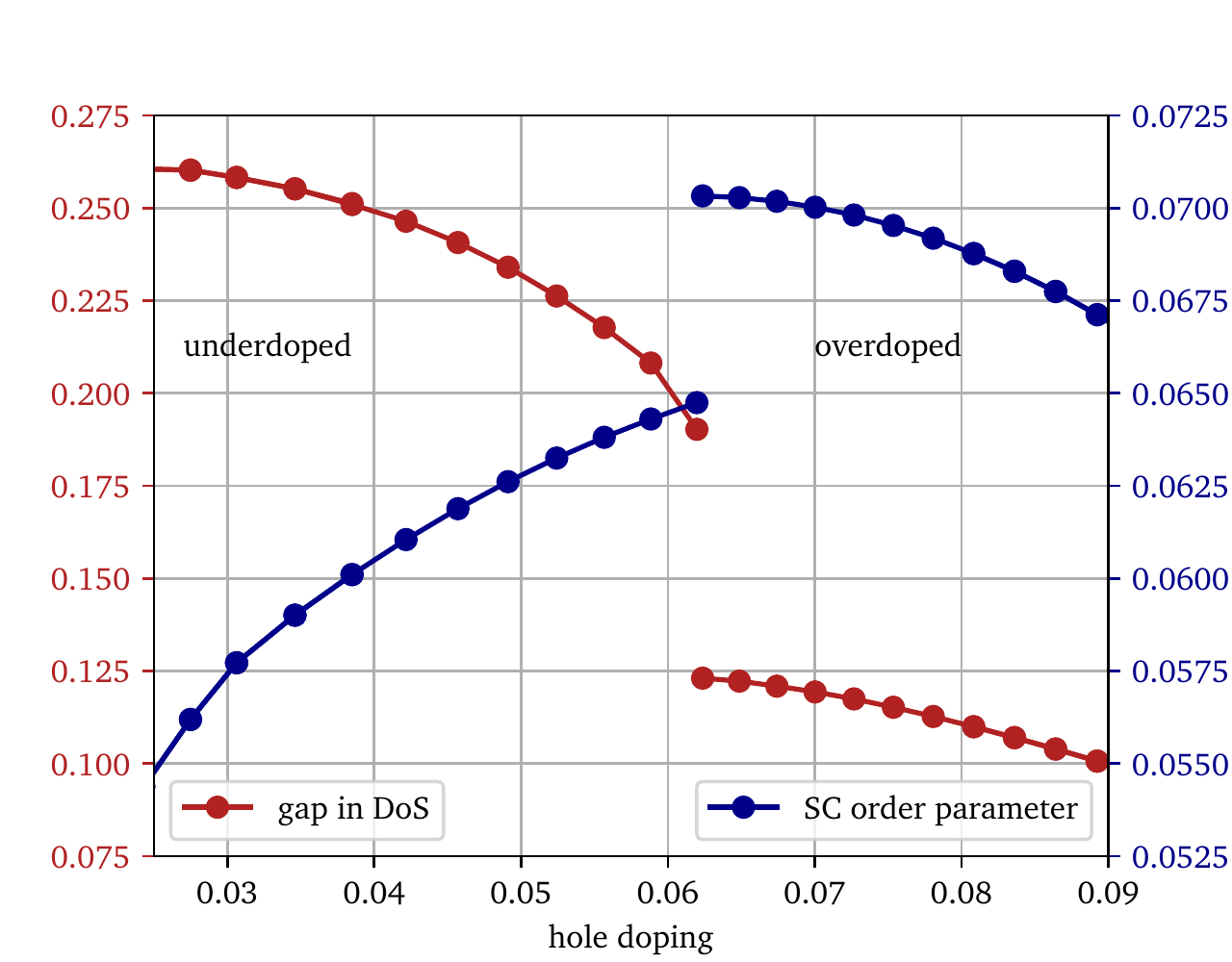}
\caption{Momentum-integrated spectral gap (red curve) and $d$-wave order parameter (blue curve) within
the SC state, as a function of hole doping, for Parameters \eqref{eq:fratino_params} and $U_d=12$.
The gap in the DoS $N(\omega)$ is calculated as the distance between the two points above and below the Fermi level where $d^2N(\omega)/d\omega^2=0$.}
\label{fig:gap_dsc}
\end{figure}

In order to understand the origin of the large gap in the underdoped solution and its relation with superconductivity, 
we show a plot of the magnitude of the gap in the DoS, along with the SC order parameter, 
as a function of hole doping (Fig.~\ref{fig:gap_dsc}).
In the overdoped solution, the spectral gap and the SC order parameter both increase towards zero doping.
This indicates that the primary source of the gap in this solution is $d$-wave superconductivity, although correlations 
make it wider than the mean-field gap (Fig.~\ref{fig:DoS_uncor_compare}(a)).
In contrast, the gap in the underdoped solution increases while the order parameter decreases~\cite{kancharla2008anomalous}.
Hence, it is evident that the dominant source of the gap in the underdoped solution is not $d$-wave 
superconductivity but something else.

%~~~~~~~~~~~~~~~~~~~~~~~~~~~~~~~~~~~~~~~~~~~~~~~~~~~~~~~~~~~~~~~~~~~~~~~~~~~~~~~
% Fig.~5
%~~~~~~~~~~~~~~~~~~~~~~~~~~~~~~~~~~~~~~~~~~~~~~~~~~~~~~~~~~~~~~~~~~~~~~~~~~~~~~~

Let us place our observations in the context of the phenomenology of cuprates.
The onset of the pseudogap in hole-doped cuprates is marked by the $T^*$ line~\cite{ding1996spectroscopic}.
It was long thought that the SC gap might emerge from the normal state pseudogap~\cite{norman2005pseudogap}.
However, it has been shown that, in the normal state, the $T^*$ line ends within the SC dome at a 
point called $p^*$~\cite{tallon2001doping,daou2009linear, collignon2017fermi}, indicating that superconductivity and the pseudogap are not directly related.
Within this context, the first-order transition between the overdoped and underdoped solutions could be understood in terms of a 
pseudogap transition. This could explain the increase of the gap magnitude even when the order parameter is 
decreasing in the underdoped solution (Fig.~\ref{fig:gap_dsc}), and also the large asymmetric gap 
(Fig.~\ref{fig:DoS_uncor_compare}(b)).
Within the normal state, the onset of the pseudogap leads to the destruction of the Fermi surface into Fermi arcs, essentially 
partially gapping the Fermi surface~\cite{timusk1999pseudogap} starting from the antinodal region in the Brillouin zone.
Within a SC state, where we already have a $d$-wave gap~\cite{loeser1996excitation,ding1996spectroscopic}, it is 
difficult to find a signature of the onset of the pseudogap in $A(\kv, 0)$. Hence, as discussed earlier, 
we look at the momentum-integrated gap to 
find the signature of the pseudogap.%giving rise to a $d$-wave gap similar to the superconducting gap.
%Since our results are in the superconducting state, looking at the spectral resolution of the gap will not let us distinguish the origin of the gap. Although Fig.~\ref{fig:gap_dsc} asserts that the origin of the large gap is not due to superconductivity, it would be helpful to look at the normal state before concluding that it is due to the pseudogap.\\

The pseudogap is a normal state property, and we indeed see its 
signatures in the normal state CDMFT solutions as well, 
namely the appearance of Fermi arcs, 
for $U_d=12$ and Parameters \eqref{eq:fratino_params} (not shown). 
The normal state CDMFT solutions, obtained with ED as the 
impurity solver, have a limitation that there is a first-order 
transition corresponding to a particle number change in the impurity model. 
Therefore, although we see a first-order transition corresponding 
to the onset of the pseudogap in the normal state, we cannot definitely connect 
it to the first-order transition in the SC state, since the transition 
in our normal state is also accompanied by a particle number 
change in the impurity model. 

\begin{figure}
	\centering
\includegraphics[trim=0.0cm 0.0cm 1.3cm 1.3cm, clip,scale=0.6]{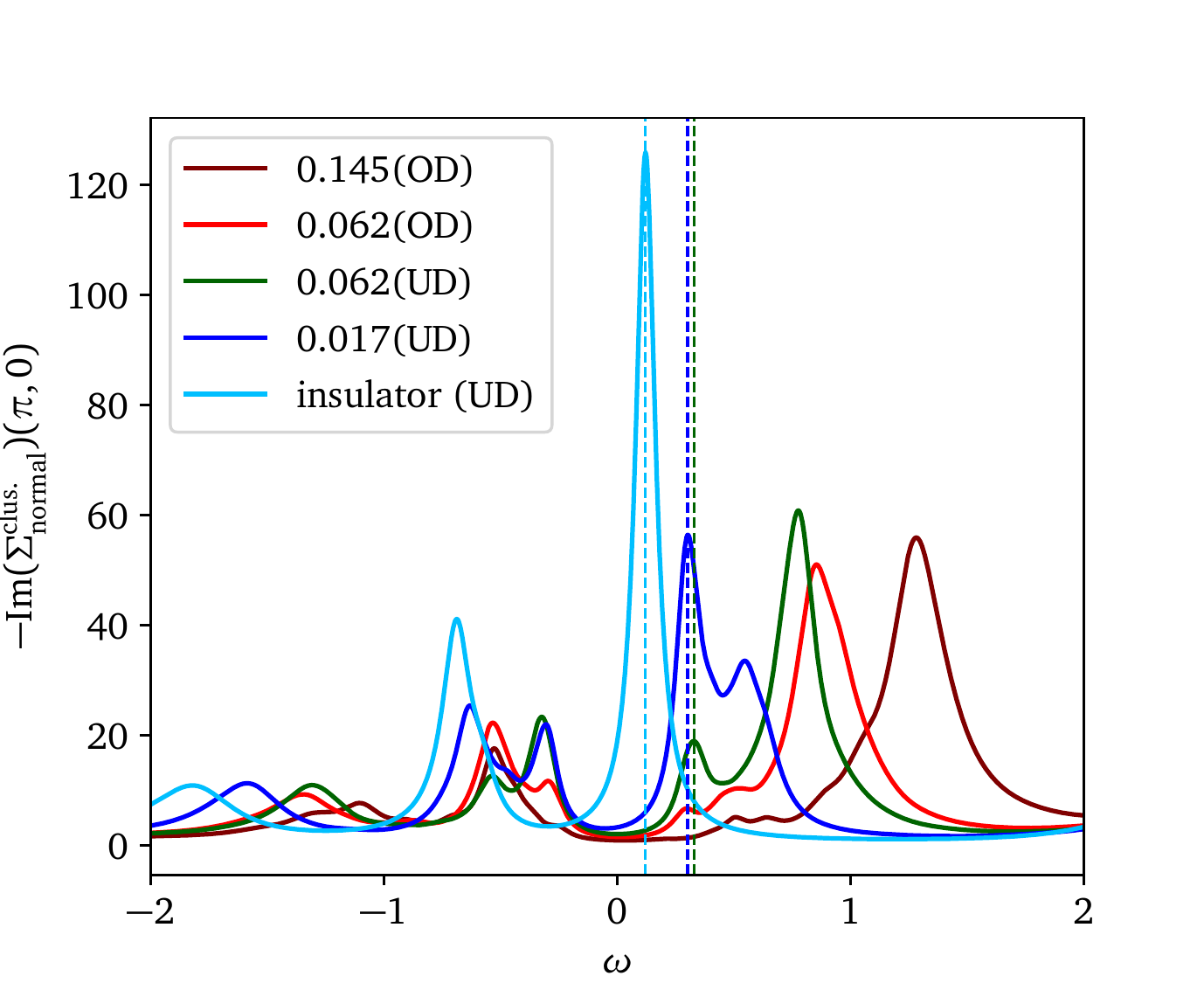}%left lower right upper 
\caption{The imaginary part of the normal component of the cluster self-energy 
is shown at the anti-nodal momentum ($\kv=(\pi, 0)$) as a function of 
frequency $\omega$, for several values of doping.
Parameters \eqref{eq:fratino_params} are used and $U_d=12$. 
The underdoped curves (marked as UD) show a peak corresponding to the 
pole in the self-energy close to the Fermi level (marked by dotted lines), 
which is absent from the overdoped curves (marked as OD).}
\label{fig:self_freq}
\end{figure}

Hence, we instead look at the normal component of the cluster 
self-energy at the antinode (Fig.~\ref{fig:self_freq}) 
to look for a signature of the pseudogap.  
The first order transition seen in Fig.~\ref{fig:d_wave_sc}
occurs between the red and green curves in Fig.~\ref{fig:self_freq} (they have the same doping 
value because of a hysteresis between the overdoped 
and underdoped solutions). 
The green curve, which 
belongs to the underdoped solution, displays a peak (marked by the green dotted line), 
absent from the overdoped solution (red and maroon curves).
This peak grows with underdoping and leads to the insulating gap at zero doping (lightblue curve).
The pseudogap is generated by such a pole close to the Fermi level as discussed in Refs~\cite{kyung2006pseudogap,sakai2016hidden,stanescu2006fermi}. 
This is also consistent with the fact that the pseudogap originates 
from Mott physics~\cite{sordi2010finite,stanescu2003pseudogap}, and hence 
strengthens our interpretation of the underdoped solution as the pseudogap state. 

%~~~~~~~~~~~~~~~~~~~~~~~~~~~~~~~~~~~~~~~~~~~~~~~~~~~~~~~~~~~~~~~~~~~~~~~~~~~~~~~
% Fig.~6
%~~~~~~~~~~~~~~~~~~~~~~~~~~~~~~~~~~~~~~~~~~~~~~~~~~~~~~~~~~~~~~~~~~~~~~~~~~~~~~~
	
\begin{figure}[htbp!]
\centering
\includegraphics[trim=0.0cm 0.0cm 1.2cm 1.2cm, clip,scale=0.7]{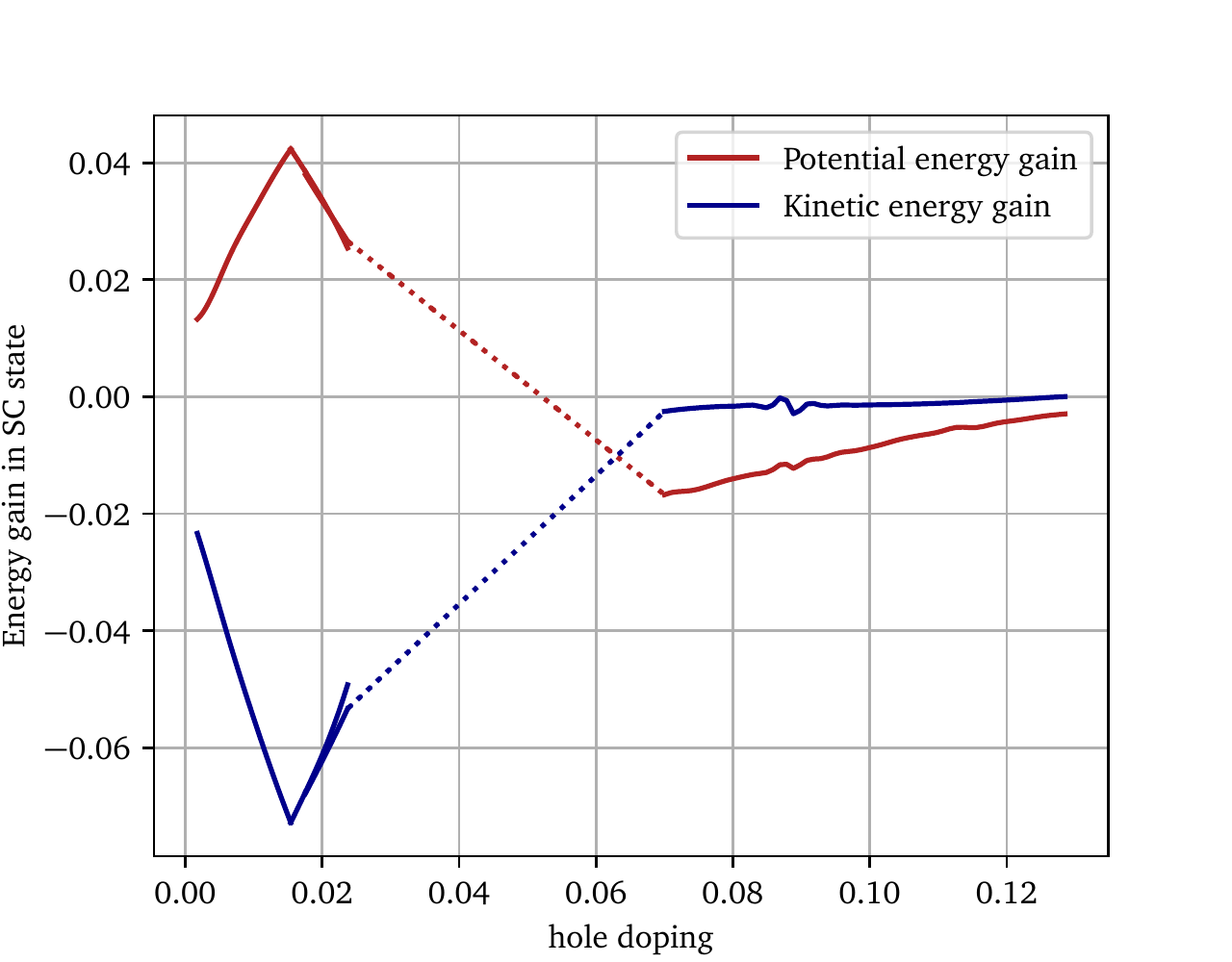}
\caption{Differences in kinetic and potential energies between the superconducting state and the normal state as a 
function of doping for Parameters \eqref{eq:fratino_params} and $U_d=12$. Superconductivity in underdoped and 
overdoped solutions is driven by kinetic energy and potential energy respectively. The normal state CDMFT 
solution is obtained with conserved particle number and spin, as opposed to the 
superconducting state which is obtained with conserved spin only. The normal state sees a transition corresponding 
to a particle number change accompanied by a jump in hole doping, hence we do not have the normal state 
solutions for a range of hole doping. We use a linear interpolation of the energy differences in that 
region, shown here by dotted lines.}
\label{fig:energy_diff}
\end{figure}

%The pseudogap transition is often understood in terms of the BCS-BEC like crossover~\cite{norman2005pseudogap,garg2005bcs}.
It is known that the nature of superconductivity changes from being driven by potential energy in the weakly 
interacting limit, to being driven by kinetic energy in the strong interaction limit.
Previous studies have reported this crossover as a function of interaction across the Mott transition~\cite{yokoyama2012crossover},  
as well as of hole doping~\cite{fratino2016organizing} across the finite-doping Mott transition~\cite{sordi2011mott}.
Figure~\ref{fig:energy_diff} shows the potential and kinetic energy gains in our CDMFT SC solutions 
(relative to the normal state CDMFT solutions) as a function of doping.
It shows that superconductivity is stabilized mostly 
by potential energy in the overdoped region, as in BCS, and by kinetic energy in the underdoped region. 
This indicates the effect of strong correlation physics in the underdoped solution and is consistent with 
the interpretation of the underdoped solution as the pseudogap state~\cite{sordi2010finite,sordi2011mott}.   
%The transition we observe between the underdoped and overdoped solutions clearly matches the above interpretation and 
%confirms that the two SC solutions are fundamentally different. 

Additionally, we observe a second, small hysteresis in the underdoped solutions for $U_d=12$ and Parameters 
\eqref{eq:fratino_params} as well as for $U_d=10,12$ and Parameters \eqref{eq:weber_params} (Fig.~\ref{fig:d_wave_sc}). 
We do not see any qualitative physical consequences of this, except for a small readjustment of the average values; 
we assume this to be just an effect of the discrete bath and hence consider it to be unphysical. 
The effect of this small hysteresis can also be seen in the condensation energy (Fig.~\ref{fig:energy_diff})
at 2\% doping, but it does not change the physics of the solutions, i.e., the source of the condensation energy. 
This reinforces the fact that such an effect can be considered as an unphysical artefact of the method. %, contrary to the above-mentioned hysteresis between the overdoped and underdoped solutions, which, as we will show below, has clear physical signatures.

%~~~~~~~~~~~~~~~~~~~~~~~~~~~~~~~~~~~~~~~~~~~~~~~~~~~~~~~~~~~~~~~~~~~~~~~~~~~~~~~
% Fig.~7
%~~~~~~~~~~~~~~~~~~~~~~~~~~~~~~~~~~~~~~~~~~~~~~~~~~~~~~~~~~~~~~~~~~~~~~~~~~~~~~~

\begin{figure}
\centering
\includegraphics[width=0.5\textwidth]{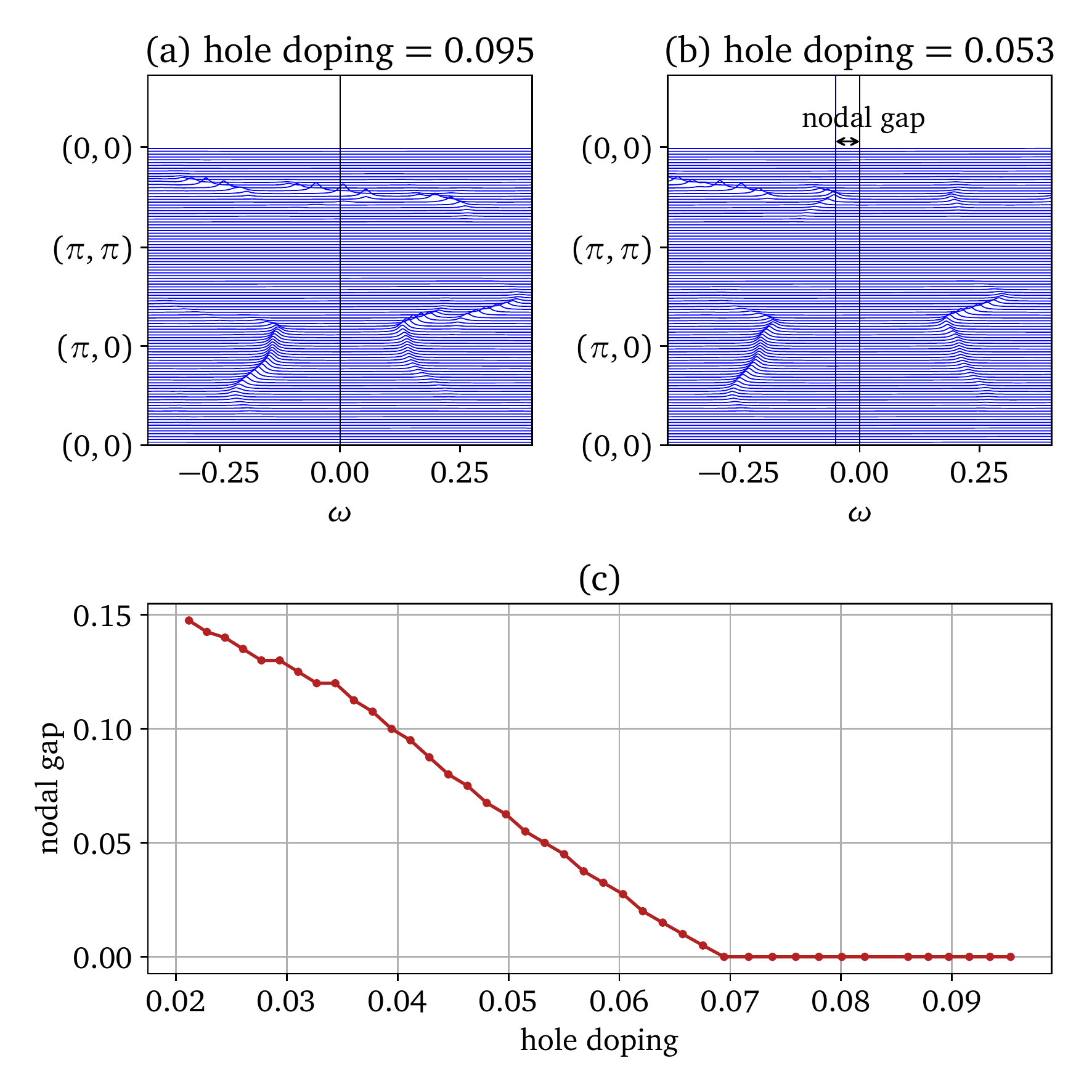}
\caption{Spectral function $A(\kv,\omega)$ along the path $(0,0)\rightarrow(\pi,0)\rightarrow(\pi, \pi)\rightarrow(0,0)$ in the Brillouin zone at $U_d = 10$, for the SC state with Parameters \eqref{eq:weber_params} (a) with node and (b) without node. (c) Gap at the node as a function of hole doping.
The spectral function is obtained from the periodized Green function~\protect\cite{senechal2000spectral}.
The nodal gap is measured from the Fermi level to the point below it where the slope of $A(\kv,\omega)$ first vanishes, i.e., at a peak.}
\label{fig:nodeless_spectrum}
\end{figure}

%It is observed that the normal state spectral function at a low value of hole doping (Fig.~\ref{spec_normal}) vanishes all along the path $(\pi,0)$ to $(\pi,\pi)$ in the Brillouin zone, whereas it takes finite values along the path $(0,0)$ to $(\pi,\pi)$. This indicates that the Fermi surface is not a full surface but a Fermi arc, validating the existence of the pseudogap phase. Hence, it seems reasonable to conclude that the large gap (Fig.~\ref{mean_field_dos_comparison_ud}) in the underdoped regime of the superconducting solutions is due to the pseudogap.\\
Another interesting feature of our results is the gradual disappearance of the nodes
in the zero-frequency spectral function $A(\kv,0)$ along the diagonal direction.
These nodes, a hallmark of $d$-wave superconductivity, disappear at a doping lower than the pseudogap transition (Fig.~\ref{fig:nodeless_spectrum}), while the SC order parameter is still finite, leading to an unusual scenario of nodeless $d$-wave superconductivity.
In this regime, the low-energy DoS develops a full gap (not shown), which we understand as an effect of strong correlations gapping the quasiparticles within the SC state.
It occurs at a doping higher than the unphysical hysteresis in Fig.~\ref{fig:d_wave_sc}, hence we think it is not related to that.
Such a nodeless superconducting regime has also been observed with ARPES~\cite{vishik2012phase}, where it is identified as a distinct phase.
However, we do not observe any signs of a sharp transition leading to this nodeless superconductivity: 
No new long-range order appears across this transition. Rather, it appears as a continuous change.

%~~~~~~~~~~~~~~~~~~~~~~~~~~~~~~~~~~~~~~~~~~~~~~~~~~~~~~~~~~~~~~~~~~~~~~~~~~~~~~~
% Fig.~8
%~~~~~~~~~~~~~~~~~~~~~~~~~~~~~~~~~~~~~~~~~~~~~~~~~~~~~~~~~~~~~~~~~~~~~~~~~~~~~~~

%===============================================================================
\section{Discussion and Conclusion}\label{sec_discussion}	

Theoretical studies using the one-band Hubbard model have observed a crossover 
between the overdoped and underdoped solutions identified with different doping dependencies of the nodal 
and anti-nodal gaps~\cite{civelli2008nodal,civelli2009doping}. In our calculations with the three-band Hubbard model, 
we observe a first-order transition clearly marking the transition from the overdoped solution, in which the 
$d$-wave order parameter increases with the gap in the DoS, to the underdoped solution in which the $d$-wave order 
parameter decreases as the DoS gap increases. The onset of such a large, increasing gap, after the 
transition, on the underdoped side further indicates that this corresponds to the pseudogap (PG) transition. 
This is further accompanied by the appearance of a pole in the normal self-energy at the antinodal region, 
which is known to generate the pseudogap~\cite{kyung2006pseudogap,sakai2016hidden}. 
Thus, the three-band model captures a richer correlation physics than the one-band model.    

The experimental doping value at which the PG ends at zero temperature, $p^*\approx 0.2$ is much higher 
than the doping at which we observe the PG transition.
It is even off by around $0.1$ for parameters \eqref{eq:weber_params} which correspond to Bi-2212.
In experiments, the doping values are mostly determined using the universal relation $T_c/T_c^{max}=1-82.6(p-0.16)^2$, 
which is verified to hold for most of the cuprate families~\cite{tallon1995generic}.
The origin of the mismatch between the experimental $p^*$ and the value we observe in our computations is not clear.
For example, the exact value of the band parameters can depend on the particular downfolding method used from the 
ab-initio band structure. It may also be that the value of $p^*$ is affected by the suppression of 
superconductivity in the experiments.

Also, the optimal doping observed in experiments is lower than the $p^*$ point~\cite{vishik2012phase,he2018rapid}.
Optimal doping is defined in terms of the maximum SC critical temperature $T_c$. 
At doping lower than the PG transition, the SC order parameter decreases in our CDMFT solutions.
This is in an apparent contradiction with experiments if we assume that $T_c$ somehow indicates the strength of the order parameter.
However, there seems to be no monotonic relation between $T_c$ and the order 
parameter~\cite{fratino2016pseudogap,fratino2016organizing}.

%~~~~~~~~~~~~~~~~~~~~~~~~~~~~~~~~~~~~~~~~~~~~~~~~~~~~~~~~~~~~~~~~~~~~~~~~~~~~~~~
% Fig.~8
\begin{figure}
\centering
\includegraphics[width=\hsize]{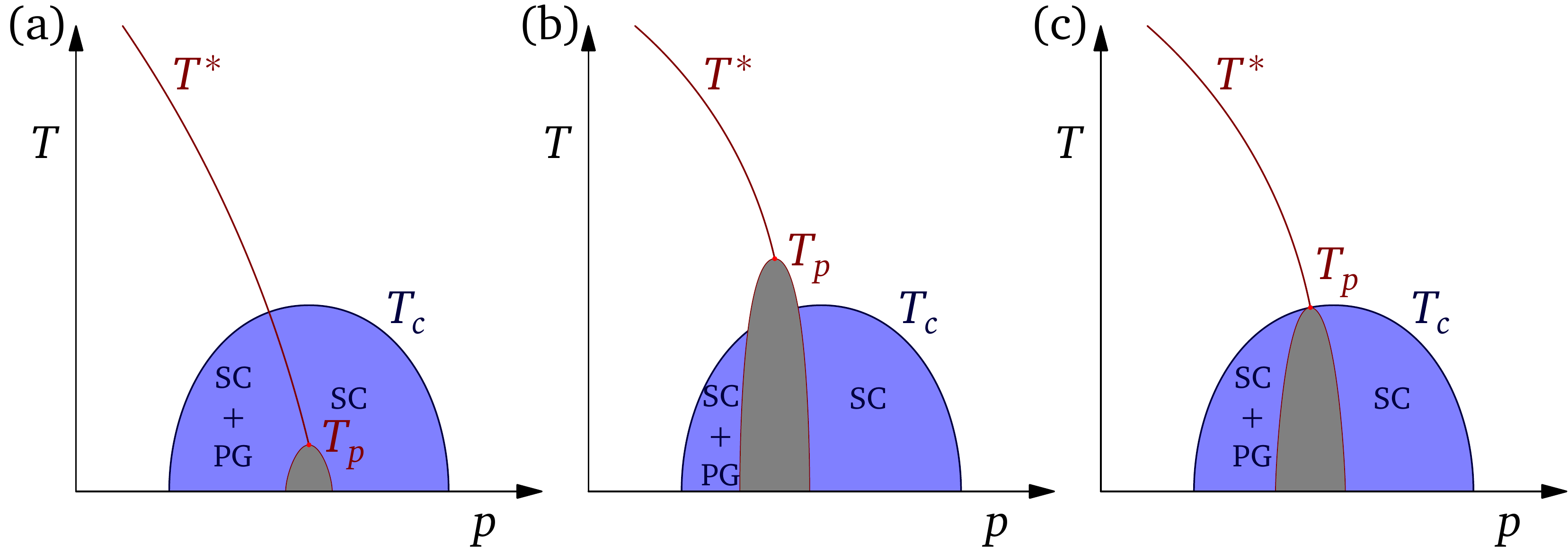}
\caption{Three scenarios are possible for the finite temperature behavior of the pseudogap (PG) transition 
depending upon the value of the PG critical point $T_p$ relative to the superconducting (SC) 
critical temperature $T_c$: (a) $T_p$ $<$ $T_c$, (b) $T_p$ $>$ $T_c$, (c) $T_p$ $=$ $T_c$ 
}
\label{fig:finite_T}
\end{figure}
%~~~~~~~~~~~~~~~~~~~~~~~~~~~~~~~~~~~~~~~~~~~~~~~~~~~~~~~~~~~~~~~~~~~~~~~~~~~~~~~

The PG transition appears as a first-order transition in the SC state at zero temperature.
We observe the transition as a jump in the value of the order parameter across a region of hole doping (represented schematically 
by the gray regions in Fig.~\ref{fig:finite_T}). 
An important question is the fate of this transition at finite temperature. 
Our first assumption is that the transition would follow the $T^*$ line since we associate it with the onset of the PG.
We can expect the transition to end at a critical point (say $T_p$), above which it exists as a crossover.
There are three possible scenarios regarding where $T_p$ lies: (a) Below $T_c$ (Fig.~\ref{fig:finite_T}(a)), 
(b) above $T_c$ (Fig.~\ref{fig:finite_T}(b)) or (c) exactly at $T_c$ (Fig.~\ref{fig:finite_T}(c)).
In the finite-temperature CDMFT study of the three-band model by Fratino~\etal~\cite{fratino2016pseudogap} 
with continuous-time quantum Monte Carlo (CT-QMC) as the impurity solver, no transition is seen within the SC phase.
Since low temperatures are difficult to reach in CT-QMC, it might be that $T_p$ lies below the lowest temperature they could probe.
It could also be that their resolution in doping is too low to observe the transition, even if $T_p$ were higher.
Moreover, they observe the finite-doping Mott transition~\cite{sordi2011mott} in the normal state, 
whose high-temperature precursor is the $T^*$ line.
It seems reasonable that the first-order transition we observe is a manifestation of the finite-doping Mott transition, 
within the SC phase.
%The finite-doping Mott transition in the work by Fratino~\etal~\cite{fratino2016pseudogap} ends at a critical point which is 
%roughly equal to the SC critical temperature $T_c$ at the same doping.
%This might indicate that the scenario of Fig.~\ref{fig:finite_T}(c) is close to the truth.
%We interpret it to be the endpoint of the $T^*$ line in the phase diagram 

One limitation of the present study is the small size of our impurity model 
(four cluster sites and eight bath sites). The spatial fluctuations in our 
calculations are restricted by the four cluster sites. And typically two 
bath sites per correlated site is considered~\cite{liebsch2011temperature} 
to adequately capture the dynamical fluctuations, hence the size of the bath 
depends on the size of the cluster. It is possible that increasing 
the size of the impurity model, which is very difficult to do at the 
moment at zero temperature, could decrease the range of the 
first-order transition; it could eventually become second-order, 
i.e., a quantum critical point, in the thermodynamic limit. However, 
a recent slave-boson calculation with the t-J model also see indication 
of a first-order pseudogap transition~\cite{mallik2018surprises}.

To summarize, we applied cluster dynamical mean field theory to the three-band Hubbard model for the cuprates at zero temperature. 
We found two distinct superconducting solutions, separated by a first-order transition as a function of hole doping. 
We interpret the underdoped solution as a manifestation of the pseudogap below $T_c$, as shown by an associated jump 
in the spectral gap, along with the appearance of a pole in the anti-nodal self-energy, and a change in the nature of the condensation energy (potential vs kinetic). 
In addition, within the underdoped solution, the $d$-wave nodes smoothly disappear very close to the insulating state.
These results are compatible with the sharp changes in the spectral gap observed as a function of doping in 
ARPES \cite{vishik2012phase,tanaka2006distinct}.
%Many experimental~\cite{tallon2001doping,daou2009linear,vishik2012phase,loret2017vertical} as well as 
%theoretical~\cite{haule2007avoided,civelli2009doping} studies have advocated for a quantum critical point, 
%emerging from a competition between the SC and PG phases.
%Our observation of a first-order transition from the SC phase to SC $+$ PG phase at zero temperature strengthens this scenario.
%Some ARPES studies~\cite{vishik2012phase,razzoli2013evolution} have further indicated the existence of an 
%additional critical point which marks the transition to a nodeless $d$-wave SC state at a doping lower than at which the PG transition occurs.
%We do observe such a scenario where the nodes disappear within the SC phase, but smoothly.
%There is no drastic change in the cluster quantities, which would signal a transition, corresponding to this.  
%Since no new long-range order appears with this nodeless SC state, we see it merely as an additional 
%manifestation of strong correlations leading to vanishing quasiparticle weights at the nodes, at a strong underdoping 
%before reaching the insulating phase.

\begin{acknowledgments}
Fruitful discussions with S.~Verret, A.-M.~Tremblay, A.~Foley, S.~Sakai, A. Mallik, S.~Chen and G.~Sordi are gratefully acknowledged.
Computational resources for this work were provided by Compute Canada and Calcul Qu\'ebec.
This work has been supported by the Natural Sciences and Engineering Research Council of Canada (NSERC) under grant RGPIN-2015-05598.
\end{acknowledgments}

%==========================================================================================
%\bibliography{references.bib}
\end{document}